# The thermal conductivity of micro/nano-porous polymers: Prediction models and applications

Haiyan Yu[1], Haochun Zhang[1, *], Jinchuan Zhao[2, 3], Jing Liu[2], Xinlin Xia[1], and Xiaohu Wu[4]

[1] *School of Energy Science and Engineering, Harbin Institute of Technology, Harbin 150001, China*
[2] *Microcellular Plastics Manufacturing Laboratory (MPML), Department of Mechanical & Industrial Engineering, University of Toronto, 5 King's College Road, Toronto, M5S 3G8, Ontario, Canada*
[3] *Cellular Polymer Science & Technology Laboratory, School of Materials Science & Engineering, Shandong University, 250061, Shandong, China*
[4] *Shandong Institute of Advanced Technology, Jinan 250100, China*

*\* Corresponding Authors: hczhang@hit.edu.cn*

## Contents

# Abstract

Micro/nano-porous polymeric material is considered a unique industrial material due to its extremely low thermal conductivity, low density, and high surface area. Therefore, it's necessary to establish an accurate thermal conductivity prediction model suiting their applicable conditions and provide a theoretical basis for expanding their applications. In this work, the development of the calculation model of equivalent thermal conductivity of micro/nano-porous polymeric materials in recent years is summarized. Firstly, it reviews the process of establishing the overall equivalent thermal conductivity calculation model for micro/nano-porous polymers. Then, the predicted calculation models of thermal conductivity are introduced separately according to the conductive and radiative thermal conductivity models. In addition, the thermal conduction part is divided into the gaseous thermal conductivity model, solid thermal conductivity model and gas-solid coupling model. Finally, it is concluded that, compared with other porous materials, there are few studies on heat transfer of micro/nano-porous polymers, especially on the particular heat transfer mechanisms such as scale effects at the micro/nanoscale. In particular, the following aspects of porous polymers still need to be further studied: micro-scaled thermal radiation, heat transfer characteristics of particular morphologies at the nanoscales, heat transfer mechanism and impact factors of micro/nano-porous polymers. Such studies would provide a more accurate prediction of thermal conductivity and a broader application in energy conversion and storage systems.

# 1. Introduction

Nowadays, as the increasing fossil fuel continuous depletion [1–3] and global severe environmental problems [4–6], exploring novel energy-saving and environmentally-friendly materials has become essential for the sustainable development of the economy and society [7–9]. The burgeoning progress of microscale and nanoscale porous polymer materials in recent years has demonstrated that this kind of materials with its low density [10–14], high surface area [15–19] and perfect thermal insulation properties [20–24] for possible challenging social and economic issues. Generally speaking, different materials, processing methods, and environmental conditions also influence the morphologies of porous polymers. Fig. 1 represents the SEM images of various micro/nano-porous polymers' morphological characteristics [24–32]. Meanwhile, the morphology of the porous polymers could be divided into open-cell and closed-cell morphologies. Open-cell porous polymers have highly interconnected seen in Fig.1 (a)-(g), while closed-cell porous polymers have isolated pores in which the gas is enclosed seen in Fig.1 (h) and (i) [33, 34].

Actually, porous polymers are widely used in various burgeoning fields, such as gas capture [35-38], catalysis [39-41], water treatment [42-44], sensors [45-47], molecular separation [48-52], food industry [53-55], proton conduction [56-58], pharmaceutical industry [59-61], drug delivery [62-64], supercapacitors [65-67], synthetic chemistry [68-70], petrochemical engineering [71-73], Li-on batteries [74-78], environmental protection [79-82], and so on [83-91].

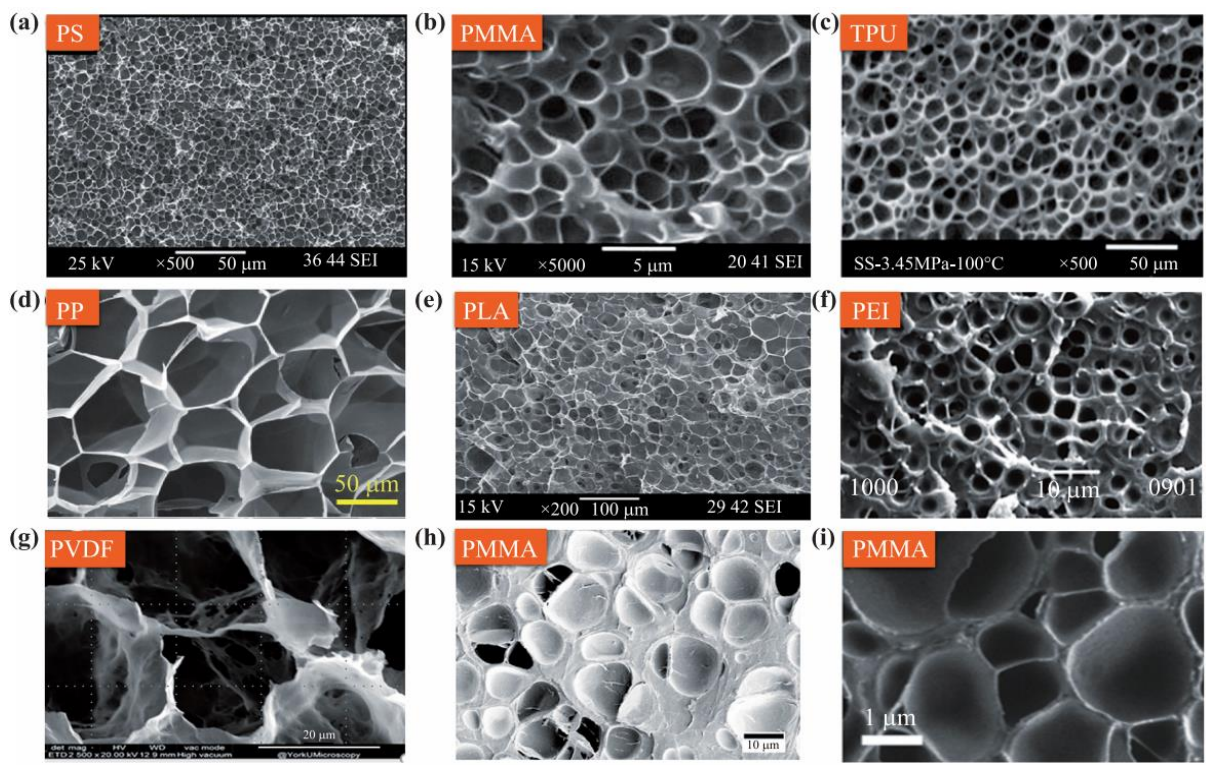

**Fig. 1.** SEM images of various micro/nano-porous polymeric materials: (a) open-cell polystyrene (PS) [25]; (b) open-cell poly(methyl methacrylate) (PMMA) [26]; (c) open-cell thermoplastic polyurethane (TPU) [27]; (d) open-cell polypropylene (PP) [24]; (e) open-cell poly(lactic acid) (PLA) [28]; (f) open-cell poly(ether imide) (PEI) [29]; (g) open-cell poly(vinylidene fluoride) (PVDF) [30]; (h) closed-cell PMMA [31]; (i) closed-cell PMMA [32].

In recent years, in order to solve the increasingly severe energy shortage problem, the potential of polymer foam in the field of energy utilization and management system has been dramatically highlighted, such as outer layer of buildings [92-97], photoenergy conversion in solar cells [98-102], electrochemical energy storage [103-108], personal thermal management devices [109-114], and so on [24, 115–119]. Therefore, it is necessary to build an accurate and appropriate thermal conductivity prediction model of micro/nano-porous polymers for better designing and applying its heat transfer characteristics [120-123] in energy conversion and storage systems [124-128], as shown in Fig. 2 [79, 122, 123, 129–133].

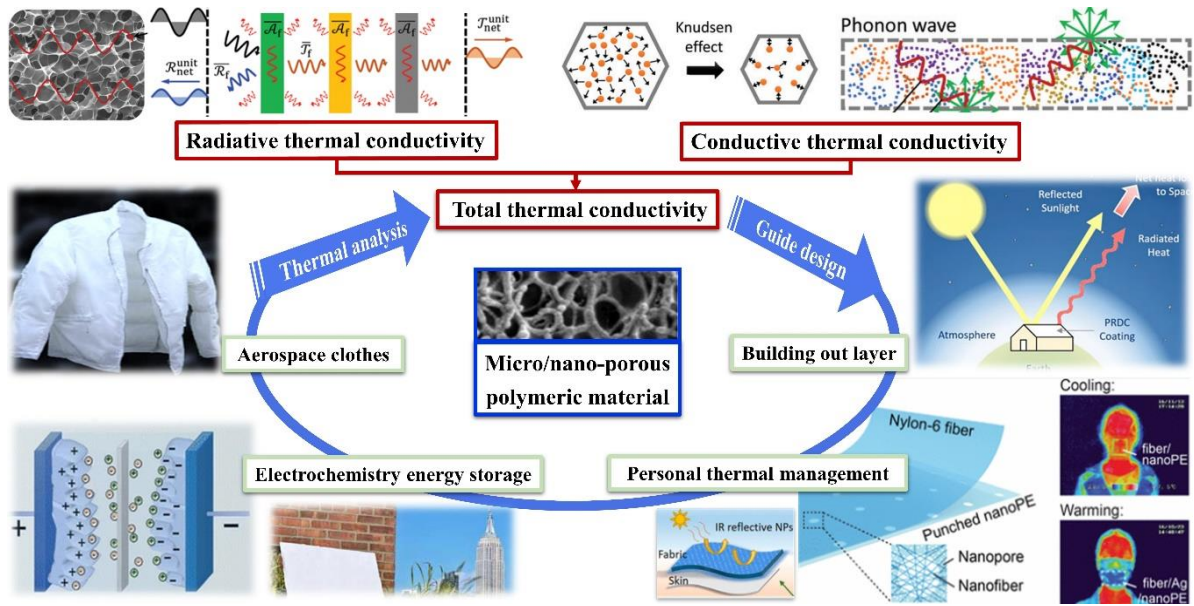

**Fig. 2.** The schematic diagram of the connection between the thermal conductivity of porous polymeric materials and their potential applications. The figure inserts are adapted from references as follows: thermal conductivity figures[122, 123], aerospace cloth figure[129], electrochemistry energy storage figure[130], personal thermal management figures[131, 132] and building out layer figures[79, 133].

This review summarizes the thermal conductivity prediction models of porous polymers to help subsequent scholars conveniently carry out their research. Firstly, it reviews the process of establishing the overall equivalent thermal conductivity calculation model for micro/nano porous polymers. Then, the thermal conductivity prediction calculation models are introduced according to thermal conduction and thermal radiation separately. In addition, the thermal conduction part is divided into the gaseous thermal conductivity calculation model, solid thermal conductivity calculation model and gas-solid coupling model. Finally, we conclude and point out the aspects that still need further study of the existing thermal conductivity prediction models of micro/nano porous polymers. Such studies would provide a more accurate prediction of thermal conductivity and a broader application in energy conversion and storage systems.

## 2. Overview of the thermal conductivity predicted models

In order to analyze the heat transfer characteristic of the porous polymers, it ought to be noted that thermal convection plays a minor role in closed-pore materials with pore sizes below 4 mm in diameter [22, 33] and in open-pore systems with pore sizes less than 2 mm [25, 134]. Therefore, in microscale and nanoscale, heat transfer of porous polymers is mainly composed of two parts: thermal conduction and thermal radiation, while thermal convection at such a confined scale is negligible [22, 25, 135]. That is, the effective total thermal conductivity of the micro/nano-porous polymers $\kappa_{total}$ is considered as composed of conductive and radiative components, as [136-139]:

$$\kappa_{total} = \kappa_{cond} + \kappa_{rad} \tag{1}$$

where the conductive thermal conductivity $\kappa_{cond}$, and the radiative thermal conductivity $\kappa_{rad}$. To sum up, the overall calculation flow chart is shown in Fig. 3. Next, the prediction models of micro/nano-porous polymers would be discussed from thermal conduction and thermal radiation parts.

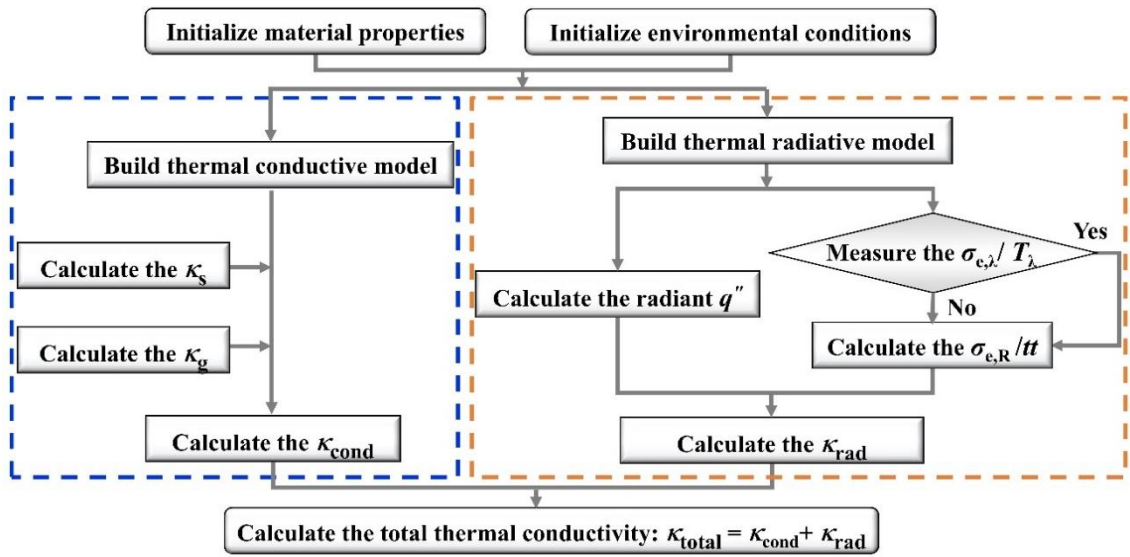

**Fig. 1.** Flow diagram of the thermal conductivity of micro/nano-porous polymers calculation progress.

## 2.1 Conductive thermal conductivity models

Generally, while establishing the conductive thermal conductivity models, the gaseous and solid thermal conductivities would be calculated separately by selected predicted models. And then bring the results into the gas-solid coupling model to solve the overall conductive thermal conductivity [14, 22, 25]. Next, these three kinds of predicted models would be introduced separately in the following sections.

## 2.1.1 Gaseous thermal conductivity models

According to the molecular kinetics theory [140], the gaseous thermal conduction is carried out by the collision of gas molecules during thermal motion [141]. The gas molecules move randomly, and the molecules at the high-temperature side have higher speed collide with molecules at the low-temperature side, causing the energy to be transferred to the lower speed molecules [141, 142].

The gaseous thermal conductivity $\kappa_g$ inside a micro/nanocellular porous polymer was determined by taking the Knudsen effect [122, 123, 143, 144] into account, which significantly led to the gaseous thermal conductivity decrease [25, 136]. This Knudsen effect implies that when cell size is equal to or smaller than the mean free path of the gas or liquid, the latter molecules collide more often with the molecules forming the surrounding solid part, reducing the energy transfer through the gas molecules [34, 143]. In a confined space, translation of gas molecules is governed by the Knudsen regime, in which the influence of cell size and mean free path on the efficiency of energy transfer is considered [123, 136, 140].

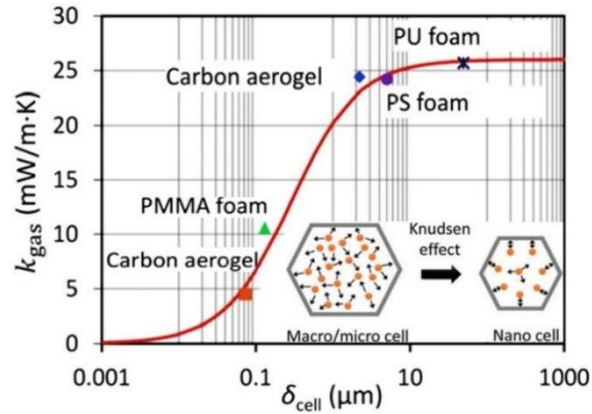

**Fig. 4.** The size effect on the gaseous thermal conductivities within the cells. The figure inserts are adapted from references as follows: Knudsen effect [122], PU foam [153], PS foam [25], Carbon aerogel [153] and PMMA foam [143].

$$k_g = \frac{k_{g,bulk}}{1+2\beta Kn} = \frac{k_{g,bulk}}{1+2\beta \frac{\Lambda}{d}} \quad (2)$$

$$2\beta = \left(\frac{2-\alpha_T}{\alpha_T}\right)\left(\frac{9\gamma-5}{\gamma+1}\right) \quad (3)$$

where $Kn = \Lambda/d$ is the Knudsen number [145–147], which refers to the ratio of the mean free path $\Lambda$ of the gas to the cell size $d$ [136]. $\beta$ is the energy interaction between gaseous molecules and the solid surface [25, 148], which is related to the thermal accommodation coefficient $\alpha_T$ [149, 150] and heat capacity ratio $\gamma$ [136, 151, 152]. The gaseous thermal conductivities are calculated for different polymer materials considering the Knudsen effect, compared with the experimental data in related reference, shown in Fig.4 [25, 122, 140, 153].

For the aerogel materials with large porosity and specific surface area, the gaseous thermal conductivity model proposed by Zeng *et al.* [154] also had a wide range of use in the prediction of aerogel models [142, 155, 156], which is simplified as:

$$\kappa_g = \frac{60.22 \times pT^{-0.5}}{0.25 S_s \rho \varphi^{-1} + 4.01 \times 10^4 \times pT^{-1}} \tag{4}$$

where $p$ is pressure, $S_s$ is specific surface area, $T$ is temperature $\varphi$ denotes porosity and $\rho$ represents the density of the porous material. Besides, considering the actual pore distribution in porous materials, Reichenauer *et al.* [153] proposed a calculation model for the gas phase thermal conductivity of porous materials based on double pore size distribution as well as the Gauss distribution. Meanwhile, Bi *et al.* [157] proposed a modified gaseous thermal conductivity model considering the randomness and inhomogeneity of pore distribution in aerogel materials.

What's more, Fricke *et al.* [158] fitted the gaseous thermal conductivity of aerogel as a function of material density based on the experimental measurement results of aerogel thermal conductivity, which can be written as:

$$\kappa_g \propto \rho^{-0.6} \tag{5}$$

It should be noted that the above equation is the contribution of the gaseous conductivity obtained directly from the experimental data to the overall thermal conductivity, not the gaseous thermal conductivity inside the pore. However, this empirical formula also needed to be refitted when the materials, densities, and boundary conditions changed.

### 2.1.2 Solid thermal conductivity models

Since the characteristic size of the porous polymer is at the microscale or even nanoscale, which is close to or smaller than the mean free path, leading to a noticeable size effect on the heat transfer progress of the solid framework [140, 141]. As a result, the solid heat transfer of polymer materials is significantly reduced [141].

In some research, the solid thermal conductivity $\kappa_s$ is theoretically extracted by considering the absence of gas in the porous polymers [25, 136], which is directly equal to the thermal conductivity of the polymer skeleton.

Considering the influence of phonon scattering at the gas-solid interface as seen in Fig.5, Wang *et al.*[122, 123] given the thermal conductivity of the thin polymer skeleton based on the kinetic theory [140, 159], as:

$$\kappa_s = (1 + \frac{\Lambda_s}{\delta})^{-1} \kappa_{bulk} \qquad (6)$$

where $\kappa_{bulk}$ is the thermal conductivity of pure solid material, $\delta$ is the characteristic size of the polymer skeleton. $\Lambda_s$ is the phonon mean free path of the bulk polymer, which can be calculated based on the thermal conductivity of bulk polymer:

$$\Lambda_s = \frac{3\kappa_{bulk}}{C_v v_g} \qquad (7)$$

where $C_v$ is the specific heat per unit volume contributed by phonons, and $v_g$ is the mean group velocity of phonons [140, 159]. Besides, this kind of modified model, considering the size effect, also wildly used the thermal conductivity calculation of aerogel materials [141, 142, 160]. Therefore, the thermal conductivity of the PS film based on Eq. (6) was calculated with a phonon mean-free path of 1.5 nm, which is in good agreement with those in the literature, as shown in Fig. 5.

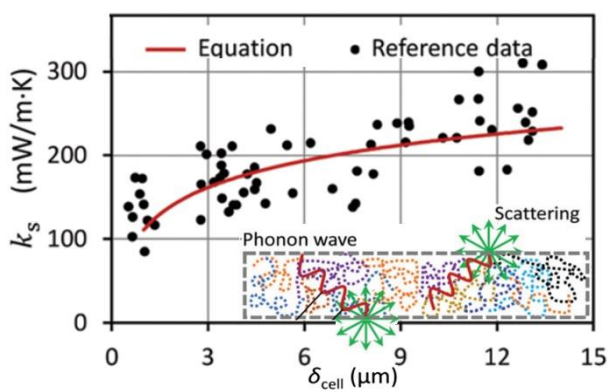

**Fig. 5.** The size effect on the solid thermal conductivities within the cells. The figure inserts are adapted from references as follows: phonon wave scattering [122] and reference data [25].

Due to the size effect, the polymer film's thermal conductivity reduces significantly with its thickness in the given range[140, 159]. What's more, Fricke *et al.* [157] fitted the solid thermal conductivity of aerogel as a function of material density based on the experimental measurement results of aerogel thermal conductivity, which can be written as:

$$\kappa_s \propto \rho^{1.5} \qquad (8)$$

It should be noted that this formula also needed to be refitted when the materials, densities, and boundary conditions changed.

### 2.1.3 Gas-solid coupling models

The existing models for the conductive thermal conductivity of porous polymers generally fall into four categories: (i) Empirical model; (ii) Equivalent circuit method; (iii) Equivalent fractal method; (iv) Numerical simulation method. In the following sections, we will introduce the simplified assumptions and application scope of these four methods.

*Empirical models*

Fricke *et al.* [157] fitted the thermal conductivity of aerogel as a function of material density based on the experimental measurement results. This method is used to adding the contribution values of thermal conductivity, which is simple in form and convenient to handle, as [142, 158]:

$$\kappa_{cond} = \kappa_g + \kappa_s \tag{9}$$

Alshrah *et al.* [161] identified the morphological features in an organic resorcinol-formaldehyde (RF) aerogel. They also correlated each mode of the heat transfer, assuming no coupling effect occurred within the solid-gas phase.

In subsequent studies, some scholars still use this method to calculate aerogel's overall equivalent thermal conductivity, such as Reichenauer *et al.*[153], Lee *et al.* [162], Deng *et al.* [163], Spagnol, *et al.* [164], Swimm *et al.* [165]. In addition, combined with the effective medium method, this empirical method [141, 166, 167] is often used to predict the equivalent thermal conductivity of polymer foam composites doped with carbon and other materials. However, this empirical prediction method usually has several empirical constants with no specific physical meaning in its equation, which is often changed with working conditions [142]. Therefore, this kind of method needs empirical correlation comparing with experimental data each time, which is specifically dependent on the materials' properties and cellular structure.

*Equivalent circuit method*

Based on Kirchhoff's law, Equivalent circuit method simplifies the complex structure into a regularized model and then connects the thermal resistance of each component in series and parallel. In the calculation of conductive thermal conductivity, this method is used as an analytical expression for the conductive thermal conductivity of the entire porous structure, determined by analyzing a network of thermal resistances for an ideal periodic unit consisting of cell struts and/or cell walls [165, 168, 169].

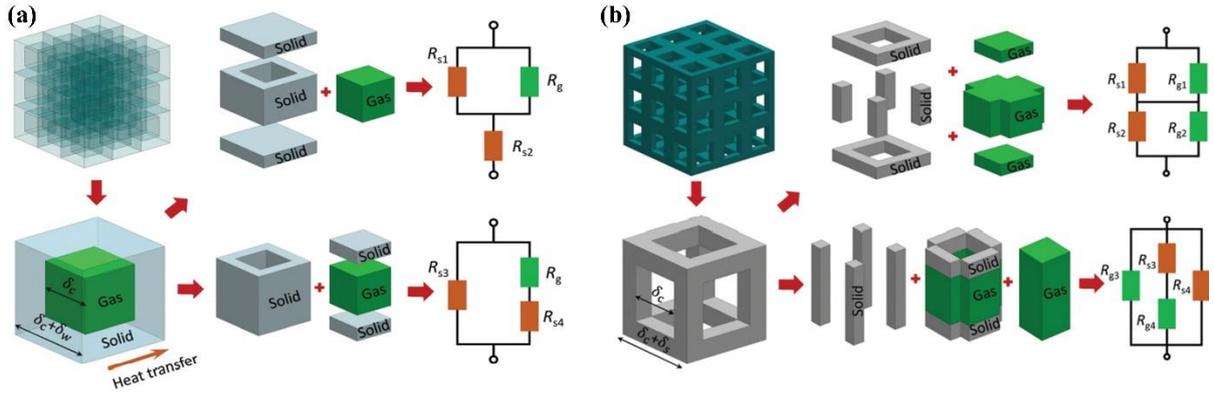

**Fig. 6.** Schematic diagram of the procedure used for deriving thermal conduction of the foam fully composed with the struts [122, 123]. (a) closed-cell model; (b) open-cell model.

Wang *et al.* [122, 123] proposed the equivalent circuit conductive thermal conductivity models of both closed-cell and open-cell porous polymers, taking the phonon scattering effect into account, as shown in Fig.6. This method has given mathematical expressions of the conductive thermal conductivity consist of the void fraction ($VF$), the strut fraction ($F_{strut}$) and the gas-to-solid thermal conductivity ratio ($k_g/k_s$). The conductive thermal conductivities of cell walls $k_{c,wall}$ and struts $k_{c,strut}$ were determined by analyzing the thermal resistance circuit of heat conduction based on the parallel-series and series-parallel methods for cubic foams [122, 123], as:

$$k_{cond} = (1 - F_{strut})k_{c,wall} + F_{strut}k_{c,strut} \tag{10}$$

$$k_{c,wall} = \frac{k_{c,wall}^{(parallel)} + k_{c,wall}^{(series)}}{2} \tag{11}$$

$$k_{c,strut} = \frac{k_{c,strut}^{(parallel)} + k_{c,strut}^{(series)}}{2} \tag{12}$$

where $k_{c,wall}^{(parallel)}$ and $k_{c,wall}^{(series)}$ are the conductive thermal conductivities of cell walls based on parallel-series and series-parallel methods, and $k_{c,strut}^{(parallel)}$ and $k_{c,strut}^{(series)}$ are the conductive thermal conductivities of struts based on parallel-series and series-parallel methods, respectively [136].

Based on this, Buahom *et al.* [136] introduced a series of alternative mathematical expressions for the conductive thermal conductivity in terms of the void fraction and the gas-to-solid thermal conductivity, as summarized in Table 1. These expressions are derived from those existing expressions based on heat conduction through closed-cell and open-cell foams.

**Table 1** Mathematical expressions of the conductive thermal conductivity models for the parallel-series and series-parallel cell walls and struts [136].

| Part | Model | Mathematical expression | |
|---|---|---|---|
| Cell walls | Parallel-Series | $\dfrac{k_{c,\text{wall}}^{(\text{parallel})}}{k_s} = \dfrac{1-\left(1-\frac{k_g}{k_s}\right)VF^{2/3}}{1-\left(1-\frac{k_g}{k_s}\right)(VF^{2/3}-VF)}$ | (13) |
| | Series-Parallel | $\dfrac{k_{c,\text{wall}}^{(\text{series})}}{k_s} = \left(1 - VF^{2/3}\right) + \dfrac{\frac{k_g}{k_s}VF^{2/3}}{\frac{k_g}{k_s}+\left(1-\frac{k_g}{k_s}\right)VF^{1/3}}$ | (14) |
| Struts | Parallel-Series | $\dfrac{k_{c,\text{strut}}^{(\text{parallel})}}{k_s} = t_{VF}^2 + \dfrac{k_g}{k_s}(1-t_{VF})^2 + \dfrac{2t_{VF}(1-t_{VF})\frac{k_g}{k_s}}{1-t_{VF}\left(1-\frac{k_g}{k_s}\right)}$ | (15) |
| | Series-Parallel | $\dfrac{k_{c,\text{strut}}^{(\text{series})}}{k_s} = \dfrac{1}{\dfrac{1-t_{VF}}{t_{VF}^2\left(1-\frac{k_g}{k_s}\right)+\frac{k_g}{k_s}}+\dfrac{t_{VF}}{1-(1-t_{VF})^2\left(1-\frac{k_g}{k_s}\right)}}$ | (16) |

where $t_{VF} = t_S/d$, which is the ratio of the thickness of rectangular struts in the open-cell cubic foams $t_S$ to the cell size $d$, can be obtained by solving $2t_{VF}^3 - 3t_{VF}^2 + 1 - VF = 0$.

Zeng *et al.* [154] and Wei *et al.* [168] also use the equivalent circuit method to calculate the overall conductive thermal conductivity of aerogel materials, while Lu *et al.* [155] using this method to calculate the conductive thermal conductivity of aerogel composite materials.

## *Equivalent fractal method*

Fractal theory can describe the disordered and random generation process of numerous porous materials at the micro and nano scales. These porous materials are different from the Euclidean description of proportional scale [170] but satisfy the fractal characteristics [171] for their self-similarity property [172,173], which represents it can be simplified the whole process by repeating unit. Based on these, the equivalent fractal method is a numerical solution [174, 175] for the conductive thermal conductivity is based on fractal geometry theory [170, 176], which presents a random structure and is independent of the scale.

The heat transfer properties of microscale and nanoscale porous materials were studied based on these dispersed fractal models, such as the Sierpinski gasket model [164, 177, 178], Vicsek model [179, 180], the Von Koch snowflakes model [181, 182] and the fractal-intersecting sphere model [141, 142]. To better represent the structure of the porous polymeric material, the fractal-intersecting sphere model was proposed by Xie *et al.* [141, 142] to calculate the conductive thermal conductivity of aerogel, as shown in Fig. 7.

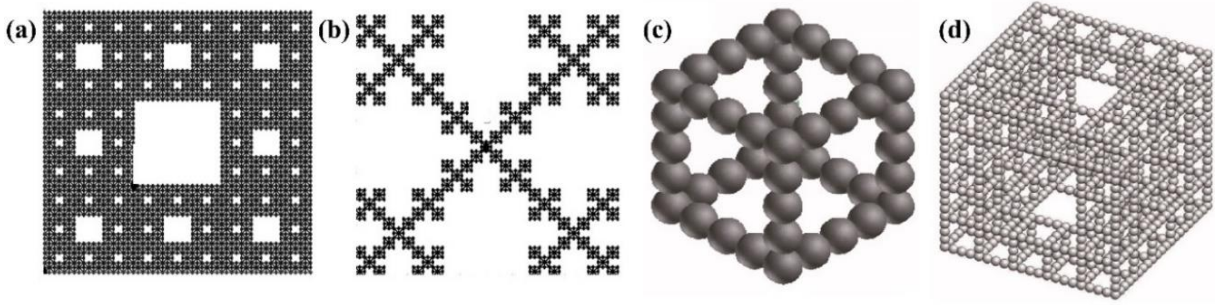

**Fig. 7.** Fractal heat transfer models: (a) Sierpinski carpet [179]; (b) Vicsek model [179]; (c) first stage Fractal-intersecting sphere model [142]; (d) second stage Fractal-intersecting sphere model [142].

Combined the equivalent circuit method and geometric of the fractal-intersecting sphere model, the conductive thermal conductivity of the fractal- intersecting model at the first stage $\kappa_{\text{cond}}^1$ and second stage $\kappa_{\text{cond}}^2$ respectively was calculated as follows [141, 142]:

$$\kappa_{\text{cond}}^1 = 4\kappa_{\text{unit}}\left(\frac{1-\xi}{2}\right)^2 + \frac{4\kappa_{\text{unit}}\kappa_g \xi(1-\xi)}{2\kappa_g \xi(1-\xi)+2\xi} + \kappa_g \xi^2 \tag{17}$$

$$\kappa_{\text{cond}}^2 = 4\kappa_{\text{cond}}^1\left(\frac{1-\xi}{2}\right)^2 + \frac{4\kappa_{\text{cond}}^1\kappa_g \xi(1-\xi)}{2\kappa_g \xi(1-\xi)+2\xi} + \kappa_g \xi^2 \tag{18}$$

where $\kappa_{\text{unit}}$ is the conductive thermal conductivity of the primary size, obtained from [183]:

$$\kappa_{\text{unit}} = 4\kappa_s\left(\frac{1-\xi}{2}\right)^2 + \frac{4\kappa_s\kappa_g \xi(1-\xi)}{2\kappa_g \xi(1-\xi)+2\kappa_s \xi} + \kappa_g \xi^2 \tag{19}$$

and $\xi = r/d$ is the aspect ratio of the solid particle radius and the cubic skeleton structure.

## *Numerical simulation method*

Sundarram *et al.* [184] used finite element analysis (FEA) [184-186] and molecular dynamics (MD) [187-190] to study the heat transfer properties of porous polymers (in particular, PMMA and PEI). Moreover, the thermal conductivity was predicted to reduce when the pore size decreased from 1 mm to 1 nm, mainly attributed to the phonon scattering effect in the solid polymer matrix. In these researches, a variety of pore configurations could represent the micro/nano-porous structure observed in polymer foams by the FEA method. In contrast, the MD method treats individual atoms in the model as point masses that interact via established interatomic potentials [184, 191]. In addition, it has also been widely used to predict the physical properties of bulk materials [192-196].

Han *et al.* [197] and Zeng *et al.* [198] used Lattice Boltzmann method (LBM) [199-203] to calculate aerogels' overall conductive thermal conductivity. Meanwhile, the scale effect of structural parameters and the temperature dependence of the thermal conductivity of nanoparticles are also studied using this simulation method.

Although the numerical method is complicated to calculate, it can reflect the influence of various factors on thermal conductivity, and it can also consider the situation of complex structures. However, the calculations in the current literature are still concentrated on the situation of some simple structures [141]. What's more, it should be noted that not to repeat considering the gas-solid interface effect in choosing the solid thermal conductivity model and setting the boundary conditions in the gas-solid coupling model.

## 2.2 Radiative thermal conductivity models

The radiant heat transfer inside the polymer material belongs to medium radiation because the polymer material is participatory media for radiant heat transfer. When incident radiant energy enters the material's interior, the material will absorb, refract, and reflect the radiation energy. Namely, the material has attenuation effects such as absorption or scattering of radiation [141]. Compared to the relatively mature development of the conductive thermal conductivity model, a model for the radiative thermal conductivity of porous polymers is still challenging. The radiative thermal conductivity becomes significant, especially in largely expanded porous dielectric materials having higher porosity [12, 22, 204, 205] and a smaller pore size [122, 206-208]. Several existing approaches for modeling the radiative thermal conductivity are based on different assumptions on the micro- and nanoscales.

### 2.2.1 Radiative transfer equation approximation

Recently, an accurate model for the radiative thermal conductivity was introduced by solving the equivalent radiant thermal conductivity through the radiative transfer equation (RTE) for simple structures [141, 209, 210].

Assuming that in a participating medium that emits, absorbs, and scattering, at position *s*, the radiation energy transfer direction $\Omega$, according to the conservation of energy, the governing equation for the transfer of radiant energy in the medium can be derived-the radiation transfer equation, as [211, 212]:

$$\frac{dI_\lambda(s,\boldsymbol{s})}{ds} = -(\sigma_{a\lambda} + \sigma_{s\lambda})I_\lambda(s,\boldsymbol{s}) + \sigma_{a\lambda}I_{b\lambda}(s) + \frac{\sigma_{s,\lambda}}{4\pi}\int_{4\pi} I_\lambda(s,\boldsymbol{s})\Phi_\lambda(\boldsymbol{s_i},\boldsymbol{s})\,d\Omega_i \qquad (20)$$

where, $I_\lambda(s,s)$ is the spectrum radiation intensity corresponding to space position *s* and transmission direction *s*, $\sigma_{a\lambda}$ and $\sigma_{s\lambda}$ are the spectrum absorption and scattering coefficients [213, 214] in the medium, respectively, and the sum of the two is the spectrum extinction coefficient $\sigma_{e\lambda}$, $\Phi_\lambda(s_i, s)$ is the scattering phase function [215, 216]. The radiation transfer equation is integrodifferential and nonlinear in the three-dimensional radiant heat transfer progress for non-gray materials unless in some simple cases, it is difficult to obtain the exact solution of the equation [140, 141]. Therefore, it is necessary to approximate the differential-integral radiation transfer equation with certain assumptions before solving it [141].

Two-flux approximation method (also called Schuster-Schwarschild approximation) [209, 217, 218] is obtained by simplifying the integral-differential radiation transfer equation into a set of linear differential equations, assuming the radiant intensity uniformly distributed in a specific angle range. However, this method is only accurate through optical thinness media, and this method is not suitable for radiation heat transfer problems with strong anisotropic scattering [141].

For the radiation intensity can be divided into several discrete directions in the entire space, Chandrasekhar [219] developed the two-flux approximation by solving the radiation intensity in these discrete directions, known as the discrete-ordinates method [220]. Besides these methods, there are different approximate methods to solve the radiative transfer equation, such as the finite volume method [221-225], discrete transfer method [226, 227], etc.

## 2.2.2 Ignore some physical processes

When the absorption or scattering of the material is unapparent, and the related term in radiation transfer equation Eq.(20) can be ignored [141]. Likewise, the influence of the radiation term is well known in conventional and microporous materials, and this term is negligible for low-density foams (<0.2 relative densities) [34, 228]. However, the conventional models used to evaluate the radiation mechanism in porous polymers assume that the infrared radiation wavelength is smaller than the pore size. It should note that this presumption is incorrect in nano-porous polymers [34, 228].

## 2.2.3 Rosseland diffusion approximation

The diffusion approximation, known as the Rosseland diffusion approximation [209, 230], is well established for optically thick media, which is widely used to calculate the radiative thermal conductivity of porous polymers. If the medium is optically thick, the medium has a strong attenuation effect on radiation, and the transmission distance of radiant energy is very short [209]. Moreover, the radiation transfer equation can be simplified based on this assumption [209].

The radiative thermal conductivity $k_{rad}$ at a mean temperature $T_m$ is usually determined by the corresponding temperature-dependent Rosseland diffusion equation [209, 230], as:

$$k_{rad} = \frac{16\sigma_{SB}T_m^3}{3\beta_R} \tag{21}$$

where $\sigma_{SB}$ is the Stefan-Boltzmann constant (5.670367 × 10$^{-8}$ W·m$^{-2}$·K$^{-4}$) and $\beta_R$ is the Rosseland mean extinction coefficient of the foam, which indicates attenuation of the radiative energy. $\beta_R$ is a harmonic mean of the spectral extinction coefficient $\beta_{\lambda,\text{foam}}$ weighted by the temperature derivative of the blackbody spectral intensity for all wavelengths within Planck's blackbody emission spectrum [231, 232], as:

$$\frac{1}{\beta_R} = \frac{\int_0^\infty \frac{1}{\beta_{\lambda,foam}} f_\lambda(\lambda,T)d\lambda}{\int_0^\infty f_\lambda(\lambda,T)d\lambda} \tag{22}$$

where $f(\lambda,T)$ is the spectral distribution of Planck's law of blackbody emission. It indicates the fraction of the radiative energy at a wavelength $\lambda$ to the overall energy over the wavelength spectrum [122, 140], as:

$$f_\lambda(\lambda,T) = \frac{\partial e_{b,\lambda}}{\partial T} = \frac{C_1}{\lambda^5 \exp\left(\frac{C_2}{\lambda T} - 1\right)} \tag{23}$$

where $e_{b,\lambda}$ is the blackbody spectral intensity at a wavelength $\lambda$. Two auxiliary constants are defined as $C_1 = 2\pi h c^2$ and $C_2 = hc/k_B$. $c$ is the speed of light in vacuum (2.99792458 × 10$^8$ m·s$^{-1}$). $h$ is the Planck constant (6.62607015 × 10$^{-34}$ J·s) and $k_B$ is the Boltzmann constant (1.380649 × 10$^{-23}$ J·K$^{-1}$) [209, 233].

As shown in Fig.3, $\beta_{\lambda,\text{foam}}$ can be measured by experiments [232] or calculate by Mie theory [234, 235]. Accordingly, the fraction of the radiative energy reflected, absorbed and transmitted at a wavelength $\lambda$ in the radiation spectrum can be defined as the spectral reflectivity ($A_\lambda$), spectral absorptivity ($R_\lambda$) and spectral transmissivity ($T_\lambda$) of the cell wall, respectively. Because $A_{\lambda,\text{wall}} + R_{\lambda,\text{wall}} + T_{\lambda,\text{wall}} = 1$, the radiative thermal conductivity could be obtained by measuring or calculating the transmissivity $T_\lambda$, as shown in Fig.3. Therefore, Gong *et al.* [22, 25] used the transmissivity deriving the radiative thermal conductivity formula of the porous polymers. Furthermore, Wang *et al.* [122, 123] considered the interference of the micro/nano-porous polymers' electromagnetic waves and modified the radiative thermal conductivity formula based on the transmissivity formula, as shown in Fig.8. This method takes the size effect into account, leading to a more accurate thermal radiation prediction, especially in the micro/nano-porous polymers. The detail of this microscale radiation method would be exhaustively introduced in the following section.

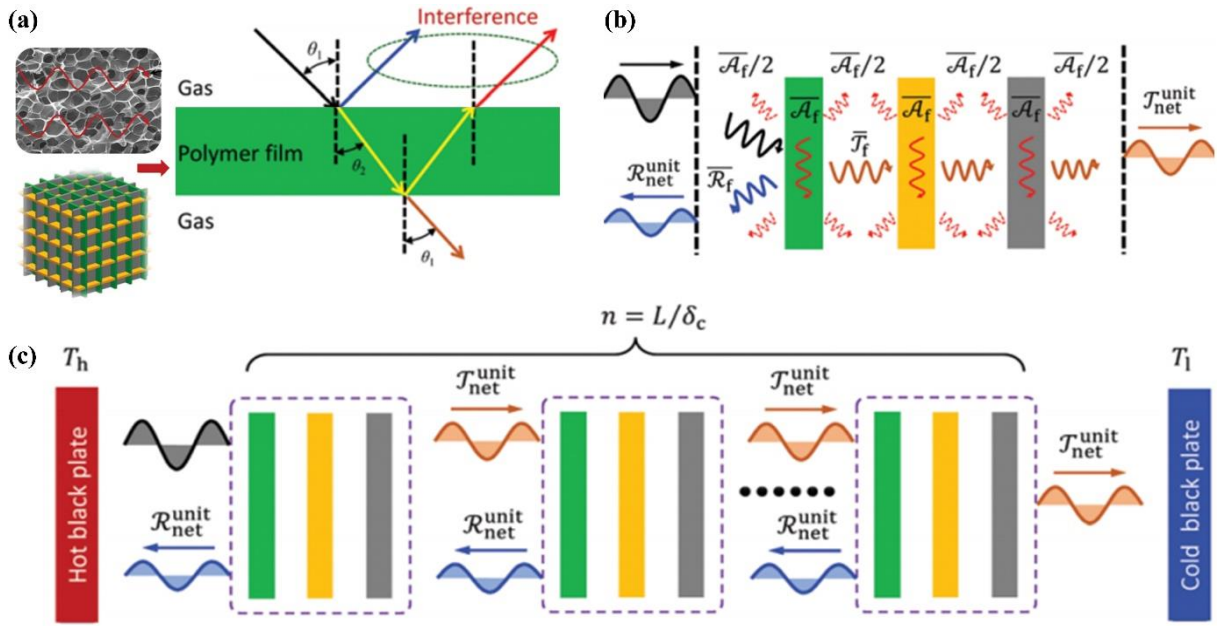

**Fig. 8.** Schematic diagrams of the thermal radiation wave propagation behaviour through a material that has multiple layers. (a) The case of a single polymer film. As the polymer film thickness is much less than the radiation wavelength, the strong thin-film interference effect can significantly reduce the reflectance. (b) The single three-slab unit is composed of a reflecting and absorbing slab and two pure absorbing slabs. Half of the whole absorbed radiant energy will be reemitted backward and the other half will be reemitted forward. (c) The case of a set of n three-slab units sandwiched between a hot black plate and a cold black plate [122].

### 2.2.4 Microscale radiation method

Since Chen *et al.* [236, 237] found that the radiant energy of two very closely located objects is much larger than predicted by the Planck Stefan-Boltzman law [237, 238], the thermal radiation process needs to consider microscale radiative heat transfer [239, 240]. Maxwell's equations describe the propagation of electromagnetic waves and their interactions with matter [160, 241]. Furthermore, the fluctuational electrodynamics combining the fluctuation-dissipation theorem with Maxwell's equations [140, 242, 243] fully describe thermal radiation's emission, propagation, and absorption in both the near and far fields [211, 244, 245].

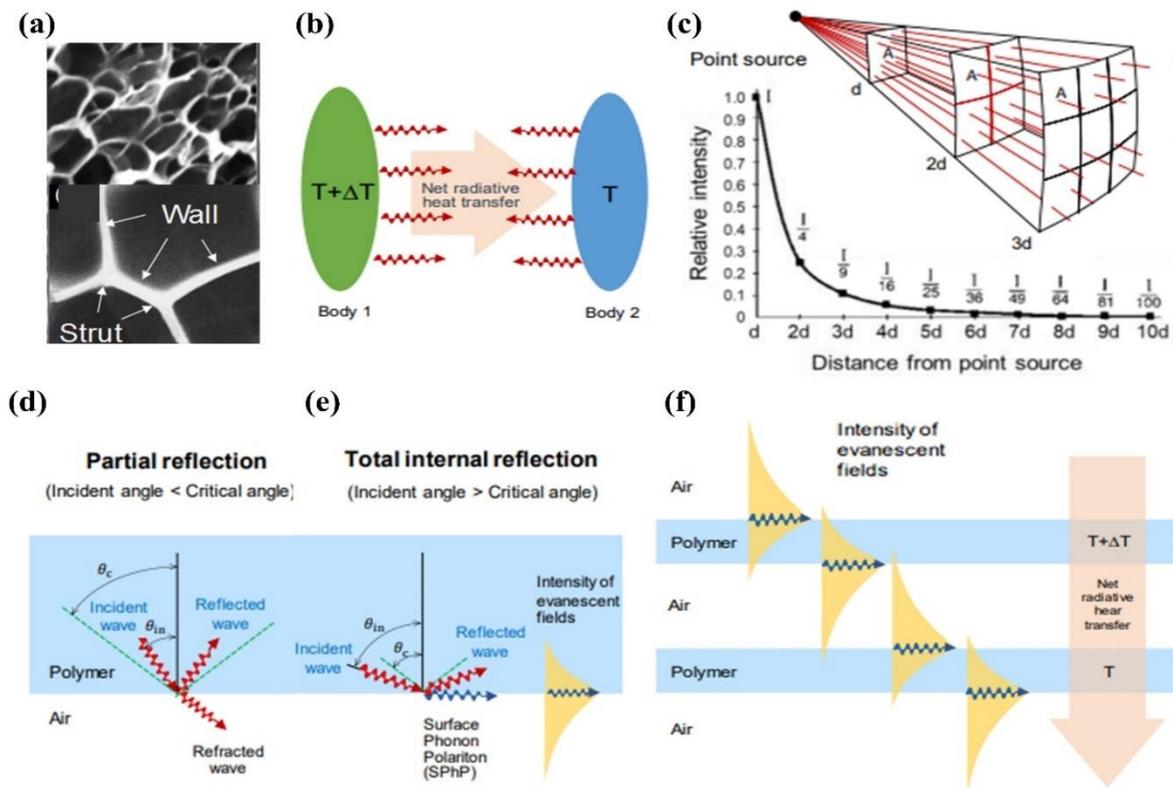

**Fig. 9.** (a) Cell walls and struts in porous polymers' morphology; (b) energy exchange between an emitter and a receiver; (c) the intensity of radiation decreases exponentially with the square of the distance from the surface due to an increase in the surface area; (d) reflection and refraction of propagating waves in partial reflection; (e) Surface phonon polaritons in polymers during total internal reflection and the intensity profile of evanescent fields; (f) tunneling of radiative energy through the foam structure by evanescent waves [136].

Buahom *et al.* [136] determined the radiative thermal conductivity was by analyzing the attenuation of radiative energy by absorption and scattering based on Mie's theory, as well as interference of propagating waves and tunneling of evanescent waves in the radiative energy.

This method illustrates that, in the porous polymer, radiation in the form of electromagnetic waves propagates in space and interacts with gas and polymer molecules [136]. Each photon carries a single discrete quantum of thermal radiation and affects the electrons in the material molecule to increase its energy level [136]. Then some of these excited electrons release the electron's excess energy and emit photons in the wavelength range according to Planck's law of blackbody emission [136, 209]. At the same time, multiple internal scattering, absorption, and re-emission occur in the porous polymer, resulting in tortuous transmission paths for radiant energy transfer, especially in such a complex structure, leading to a tortious transport path of radiative energy transfer [22, 136, 209].

As shown in Eq. (22), $\beta_{\lambda,\text{foam}}$ can be obtained by assuming randomly oriented concave-triangular struts and cell walls independently as [136]:

$$\beta_{\lambda,foam} = F_{strut}\beta_{\lambda,strut}^{(eff)} + (1-F_{strut})\beta_{\lambda,wall}^{(eff)} \qquad (24)$$

where $\beta_{\lambda,\text{strut}}^{(\text{eff})}$ is the contribution of struts and $\beta_{\lambda,\text{wall}}^{(\text{eff})}$ is the contribution of cell walls, the volume fraction of polymer located in struts, the strut fraction $F_{\text{strut}}$, is used as the weighted factor in the calculation. The effective spectral extinction coefficient of cell wall $\beta_{\lambda,wall}^{(eff)}$ can be calculated as [136]:

$$\beta_{\lambda,wall}^{(eff)} = N_{v,wall}C_{wall}^{(geo)}\int_0^{\pi/2}(1+R_{\lambda,wall}-T_{\lambda,wall})\frac{\sin(2\phi_1)}{4}d\phi_1 \qquad (25)$$

where $N_{v,\text{wall}}$ is the number of cell walls per unit volume of foam, $C_{\text{wall}}^{(\text{geo})} = C_{\lambda,\text{wall}}^{(\text{ext})}/Q_\lambda^{(\text{ext})}$, $C_{\lambda,\text{wall}}^{(\text{ext})}$ is the spectral extinction cross-section of the cell wall for the incident beam interacts with a wall tilted by the incident angle $\phi_1$. the spectral extinction efficiency factor, the ratio of the spectral extinction cross-section to the geometric cross-section of the cell wall at normal incident $Q_{\lambda,\text{wall}}^{(\text{ext})} = R_{\lambda,\text{wall}} + \frac{1}{2}A_{\lambda,\text{wall}}$. Based on Fresnel formula, $R_\lambda$ and $T_\lambda$ in Eq. (25) can be written as [246]:

$$R_{\lambda,\text{wall}} = \frac{\rho_{\lambda,\text{gs}}^2 + \rho_{\lambda,\text{sg}}^2 + 2\rho_{\lambda,\text{gs}}\rho_{\lambda,\text{sg}}\cos 2\tilde{\beta}_\lambda}{1+\rho_{\lambda,\text{gs}}^2\rho_{\lambda,\text{sg}}^2 + 2\rho_{\lambda,\text{gs}}\rho_{\lambda,\text{sg}}\cos 2\tilde{\beta}_\lambda} \quad \text{and} \quad T_{\lambda,\text{wall}} = \frac{\tau_{\lambda,\text{gs}}^2\tau_{\lambda,\text{sg}}^2}{1+\rho_{\lambda,\text{gs}}^2\rho_{\lambda,\text{sg}}^2 + 2\rho_{\lambda,\text{gs}}\rho_{\lambda,\text{sg}}\cos 2\tilde{\beta}_\lambda} \qquad (26)$$

where the spectral reflection and transmission coefficients at the gas-solid ($\rho_{\lambda,\text{gs}}$, $\tau_{\lambda,\text{gs}}$) and solid-gas ($\rho_{\lambda,\text{sg}}$, $\tau_{\lambda,\text{sg}}$) interfaces are defined as [246]:

$$\rho_{\lambda,\text{gs}} = -\rho_{\lambda,\text{gs}} = \frac{\hat{n}_{\lambda,\text{g}}\cos\phi_1 - \hat{n}_{\lambda,\text{s}}\cos\phi_2}{\hat{n}_{\lambda,\text{g}}\cos\phi_1 + \hat{n}_{\lambda,\text{s}}\cos\phi_2} \qquad (27)$$

$$\tau_{\lambda,\text{gs}} = \frac{2\hat{n}_{\lambda,\text{g}}\cos\phi_1}{\hat{n}_{\lambda,\text{g}}\cos\phi_1+\hat{n}_{\lambda,\text{s}}\cos\phi_2} \quad \text{and} \quad \tau_{\lambda,\text{sg}} = \frac{2\hat{n}_{\lambda,\text{s}}\cos\phi_2}{\hat{n}_{\lambda,\text{g}}\cos\phi_1+\hat{n}_{\lambda,\text{s}}\cos\phi_2} \qquad (28)$$

where $\hat{n}_{\lambda,s} = n_{\lambda,s} + i\kappa_{\lambda,s}$ and $\hat{n}_{\lambda,g} = n_{\lambda,g}$ are the spectral complex refractive indices of polymer film and non-absorbing gas.

The spectral extinction coefficient of the strut can be expressed in the number density of struts and the geometric cross-section [136]:

$$\beta_{\lambda,strut}^{(eff)} = N_{v,strut} C_{strut}^{(geo)} \int_0^{\pi/2} \left[ \left( Q_{\lambda,strut}^{(ext)} - Q_{\lambda,strut}^{(sca)} \right) sin^2\phi - g_{\lambda,strut} Q_{\lambda,strut}^{(sca)} cos^2\phi \right] cos\phi d\phi \quad (29)$$

where $N_{v,strut}$ is the number of struts per unit volume of porous polymers, which is dependent on the porosity, the cell size, and the strut fraction. $C_{strut}^{(geo)}$ is the normally projected geometric cross-section of a strut. $g_{\lambda,strut}$ is the asymmetry factor of the strut. $Q_{\lambda,strut}^{(sca)}$ and $Q_{\lambda,strut}^{(ext)}$ are spectral scattering and extinction efficiency factors of the strut, separately split into the TM and TE incident modes [136].

## 3. Summary and prospects

This study introduces and reviews the existing thermal conductivity calculation models of micro/nano- porous polymers. Firstly, it reviews the process of establishing the overall equivalent thermal conductivity calculation model for micro/nano- porous polymers. Then, the thermal conductivity prediction model is introduced separately based on thermal conduction and thermal radiation. In addition, the thermal conduction part is divided into the gaseous thermal conductivity calculation model, solid thermal conductivity calculation model and gas-solid coupling model. Finally, it is concluded that, compared with other porous materials, there are few studies on heat transfer of micro/nano-porous polymers, especially on the particular heat transfer mechanisms such as scale effects at the micro/nanoscale. In particular, the following aspects of porous polymers still need to be further studied:

(1) The study of characteristics and influences in micro/nano-porous polymers brought by microscale thermal radiation.

(2) The research of the heat transfer characteristics of open/ closed-pores and particular morphologies at the nanoscale.

(3) Heat transfer mechanism and impact factors of micro/nano-porous polymers. By analyzing the various influencing factors on heat transfer performance, the porous polymers can be further optimized for efficient thermal insulation materials, which are well used in energy conversion and storage systems.

## Acknowledgments

The authors are grateful to the National Natural Science Foundation of China (No.51776050 and No.51536001).